\documentclass[sn-mathphys,Numbered]{sn-jnl}

\usepackage{graphicx}%
\usepackage{multirow}%
\usepackage{amsmath,amssymb,amsfonts,bm}%
\usepackage{amsthm}%
\usepackage{tensor}%
\usepackage{mathrsfs}%
\usepackage[title]{appendix}%
\usepackage{xcolor}%
\usepackage{textcomp}%
\usepackage{manyfoot}%
\usepackage{booktabs}%
\usepackage{algorithm}%
\usepackage{algorithmicx}%
\usepackage{algpseudocode}%
\usepackage{listings}%
\usepackage{comment}

\theoremstyle{thmstyleone}%
\theoremstyle{thmstyletwo}%

\theoremstyle{thmstylethree}%

\def\doi{http://doi.org}

\newcommand{\onehalf}{{\textstyle\frac{1}{2}}}

\newcommand{\pfrac}[2]{\frac{\partial{#1}}{\partial{#2}}}

\newcommand{\ppfrac}[3]{\frac{\partial^{2}{#1}}{\partial{#2}\partial{#3}}}

\newcommand{\detpartial}[2]{\left| \pfrac{#1}{#2} \right|}
\newcommand{\eref}[1]{Eq.~(\ref{#1})}

\newcommand{\diff}[2]{\frac{\partial #1}{\partial #2}}

\newcommand{\dt}{\text{d}}

\newcommand{\Dderr}{\overrightarrow{\mathcal{D}}}
\newcommand{\Dderl}{\overleftarrow{\mathcal{D}}}

\newcommand{\must}{\stackrel{!}{=}}
\newcommand{\HO}{\Omega}
\newcommand{\ho}{\omega}

\newcommand{\hoc}{\tilde{q}}

\newcommand{\HCd}{\mathcal{H}}

\newcommand{\FCd}{\mathcal{F}}

\def\HCdt0{\tilde{\HCd}_{0}}

\newcommand{\LCd}{\mathcal{L}}
\newcommand{\dd}{\,\mathrm{d}}
\newcommand{\rmi}{\mathrm{i}}

\begin{document}

\title[Article Title]{Massive propagating modes of torsion}


\author*[1,2]{\fnm{Vladimir} \sur{Denk}}\email{denk@fias.uni-frankfurt.de}

\author[1]{\fnm{David} \sur{Vasak}}\email{vasak@fias.uni-frankfurt.de}

\author[1]{\fnm{Johannes} \sur{Kirsch}}\email{jkirsch@fias.uni-frankfurt.de}

\affil*[1]{\orgname{Frankfurt Institute for Advanced Studies (FIAS)}, \orgaddress{\street{Ruth-Moufang-Strasse 1}, \city{Frankfurt}, \postcode{60438}, \country{Germany}}}

\affil[2]{\orgdiv{Institute for Theoretical Physics}, \orgname{Goethe University}, \orgaddress{\street{Max-von-Laue-Strasse 1}, \city{Frankfurt}, \postcode{60438}, \country{Germany}}}


\abstract{The dynamics of the torsion field is analyzed in the framework of the Covariant Canonical Gauge Theory of Gravity (CCGG), a De~Donder-Weyl Hamiltonian formulation of gauge gravity. The action is quadratic in both, the torsion and the Riemann-Cartan tensor. 
Since the latter adds the derivative of torsion to the equations of motion, torsion is no longer identical to spin density, as in the Einstein-Cartan theory, but an additional propagating degree of freedom. 
As torsion turns out to be totally anti-symmetric, it can be parametrised via a single axial vector. 
It is shown in this paper that, in the weak torsion limit, the axial vector obeys a wave equation with an effective mass term which is partially dependent on the scalar curvature. 
The source of torsion is thereby given by the fermion axial current which is the net fermionic spin density of the system. 
Possible measurable effects and approaches to experimental analysis are addressed. 
For example, neutron star mergers could act as a dipoles or quadrupoles for torsional radiation, and an analysis of radiation of pulsars could lead to a detection of torsion wave background radiation.}

\maketitle

Over the years, with rising amounts of astronomical data, discrepancies between observations and the theory of gravitation and matter have led to increasing need to modify the original ansatz by Einstein and Hilbert published in 1915 \cite{Hilbert}. Albeit the latter is very accurate in describing physics on the scale of the solar system, the life-cycles of stars, gravitational waves and black holes, and is widely used for satellite navigation and GPS, it fails when applied to systems the size of galaxies or to the universe as a whole. A preliminary remedy to align observations and theory was the ad hoc introduction of dark matter and dark energy that provide a phenomenological so-called Concordance Model of the universe. Unfortunately, the underlying physical nature of both, dark matter and dark energy, is still mysterious despite of intensive research.\par
An alternative avenue to modifications of the dynamics of spacetime and matter is treating gravity as a gauge theory \`{a} la Yang-Mills as pioneered by Sciama, Utiyama and Kibble \cite{sciama53, utiyama56, kibble61} and nowadays known under the tag of Poincar\'{e} Gauge Theory \cite{Hehl17}. Here we follow a similar philosophy but rely on the rigorous formalism of the manifestly covariant transformation theory which is a well-defined methodology for implementing local symmetries in semi-classical systems of relativistic fields~\cite{struckmeier08}. That framework applied to matter fields in curved spacetimes is known as Covariant Canonical Gauge Gravity (CCGG)~\cite{cantran,Vasak:2023ncz}. Starting from the manifestly covariant De~Donder-Weyl (DW) Hamiltonian formulation, that framework unambiguously derives the coupling of matter and gravity mediated by newly introduced gauge fields \cite{struckmeier08,cantranII}. Therein the Hamiltonian formalism demands the action to be necessarily quadratic in the canonical momenta \cite{Benisty:2018ufz}, which leads to a term proportional to $R^\mu{}_{\nu\alpha\beta}R^\nu{}_\mu{}^{\alpha\beta}$ in the Lagrangian. In addition, the metric-affine (aka Palatini) structure of the spacetime fields allows for a non-symmetric connection, hence torsion of spacetime emerges as an additional degree of freedom.\par
In this paper we want to take a closer look at torsion as it arises in the CCGG formalism. Specifically, we wish to investigate small excitations that can be described by waves. 
As shown in~\cite[p. 49]{Vasak:2023tiu} and reviewed below, any torsion tensor that solves the field equations must be completely anti-symmetric and can be expressed by an axial vector. 
This puts it in close resemblance to vector (gauge) fields like photons and gluons. 
Indeed, 
a wave equation for the torsion axial vector in curved geometry will be deduced, formally similar to the wave equation for a massive Proca field.
Thereby net spin density of fermions, i.e.\ the difference between spin-up and spin-down states, will be identified as the main source of those torsion waves.  
Bosons, on the other hand, are shown not to directly interact with torsion~\cite[eq. (144) and (145)]{Vasak:2023tiu}.

\medskip
The paper is structured as follows. 
In the first section the basics of CCGG are reviewed, with details of the underlying canonical transformation theory being summarized in Appendix~\ref{appDWHtrafo}.
In the second section the field equation for the connection is analyzed and recast into the form of a wave equation for the torsion field. 
In the third section, the sources of torsion are derived. 
The paper concludes with a discussion of the key findings, and with an outlook on possible applications and emerging research topics.\par
Throughout this work the conventions are natural units $\hbar=c=1$ and the metric signature $(+---)$. The Riemann-Cartan tensor is defined as
\begin{align}
R^\alpha{}_{\beta\gamma\delta}=\partial_\gamma \Gamma^\alpha{}_{\beta\delta}-\partial_\delta \Gamma^\alpha{}_{\beta\gamma}+\Gamma^\alpha{}_{\mu\gamma}\Gamma^\mu{}_{\beta\delta}-\Gamma^\alpha{}_{\mu\delta}\Gamma^\mu{}_{\beta\gamma},
\end{align}
where $\Gamma^\alpha{}_{\beta\gamma}$ is the asymmetric affine connection that is a priori independent of the metric. Concerning the ordering of indices, the covariant derivative is given by $\nabla_\mu a^\nu=\partial_\mu a^\nu +\Gamma^\nu{}_{\alpha\mu}a^\alpha$.

\section{The CCGG formalism}
The outset of the CCGG framework is the manifestly covariant De~Donder-Weyl Hamiltonian formalism, in which the canonical momenta are not only constructed from the time derivatives of fields but by  
covariant field derivatives across all space-time coordinates.
This enables to deploy the covariant field-theoretical version of the canonical transformation theory. At its heart are the so-called generating functions enforcing invariance of systems with respect to a selected local symmetry transformations~\cite{struckmeier08}. 
Restoring that invariance requires the introduction of compensating gauge fields.  
For internal symmetry groups like U($1$) or SU($3$), that gives rise to Yang-Mills theories with vector gauge fields, photons or gluons, respectively~\cite{Struckmeier:2012de}.
In the case of gravity, the gauge group is $\text{Diff}(M)\times$SO($1,3$), and the vierbeins (also known as tetrads) and the spin connection emerge as the gauge fields for the subgroups $\text{Diff}(M)$ and SO($1,3$), respectively\footnote{
This corresponds formally to the Poincare gauge group. However, in contrast to PGT, in CCGG the diffeomorphisms are passive local chart transitions rather than active diffeomorphisms. Also the interpretation of the vierbein as a gauge field of translation~\cite{kibble61} is not necessary~\cite{utiyama56}, see also the discussion in~\cite{Vasak:2023tiu}.}.
(In order to make the paper to some degree self-contained we include details of the framework in~Appendix~\ref{appDWHtrafo} and refer to~\cite{Vasak:2023tiu}.)
\par
A key assumption of this approach is, that the underlying Hamiltonian respectively Lagrangian must be non-degenerate since otherwise the Legendre transform would not exist.  
This mathematical condition has now a significant impact on the physical content of the theory as the DW Hamiltonian must be at least quadratic in the involved canonical momentum fields~\cite{Benisty:2018ufz}.  
Consequently, the corresponding Lagrangian of the gravitational system must extend the Einstein-Hilbert ansatz by a term proportional to the square of the Riemann-Cartan tensor, $R^\mu{}_{\nu\alpha\beta}R^\nu{}_\mu{}^{\alpha\beta}$, and in theories with torsion  by the square of the torsion tensor, $S_{\alpha\beta\gamma}S^{\alpha\beta\gamma}$. \par
Following these requirements, the simplest Hamiltonian density that ``deforms'' the Einstein ansatz but retains its phenomenology on the solar system scale, is \cite[eq. 93]{Vasak:2023tiu}:
\begin{align}\label{def:HGR}
\tilde{\mathcal{H}}_\text{Gr}&=\frac{1}{4\varepsilon g_1}\tilde{q}^l{}^{m\alpha\beta}\tilde{q}_m{}_l{}_{\alpha\beta}+g_2\tilde{q}_l{}^{m\alpha\beta}e^l{}_\alpha \eta_{mn} e^n{}_\beta+\frac{1}{2 g_3 \varepsilon}\tilde{k}^{l\alpha\beta}\tilde{k}_{l\alpha\beta},
\end{align}
where $\varepsilon$ is the determinant $\varepsilon=\det(e^i{}_\alpha)$ of the vierbein facilitating an invariant volume element\footnote{Note, that there is a relative minus sign in the definition of $g_2$ with respect to \cite{Vasak:2023tiu}.}. A tilde above a symbol denotes a tensor density, e.g. $\tilde{\mathcal{H}}_\text{Gr} := \varepsilon\,{\mathcal{H}}_\text{Gr}$. The $g_i$ are free parameters of the theory, which have to be constrained through comparison with experimental data.\par
In the following the generic action integral~\eqref{def:actionintegral0} will be understood as a specific ansatz containing the DW Hamiltonian~\eqref{def:HGR}.
Since scalar and vector fields do not directly interact with torsion~\cite[p. 51]{Vasak:2023tiu}, we can safely ignore them in the action for the following weak-field limit analysis. 
The fermion field, though, does couple to the gauge fields 
and thus needs to be evaluated in detail.
Here a non-degenerate version of the Dirac Lagrangian is applied wielding quadratic "velocity" terms:
\begin{align}
\LCd_{DG}&=\frac{\rmi\ell}{3}\left(\pfrac{\bar{\psi}}{x^{\alpha}}-
\frac{\rmi}{2\ell}\bar{\psi}\,\gamma^{i}e_i{}^\alpha\right)\sigma^{kl}\,e_k{}^\alpha \,e_l{}^\beta
\left(\pfrac{\psi}{x^{\beta}}+\frac{\rmi}{2\ell}\gamma^n e_n{}^\beta \psi\right)-\left(m-\ell^{-1}\right)\bar{\psi}\psi\,.
\label{eq:ld-dirac-coupling0}
\end{align}
$\sigma^{ij}:=\frac{i}{2}[\gamma^i,\gamma^j]$, $\omega^j{}_{i\mu}$ is the spin connection and $\ell$ is an emerging length parameter\footnote{With respect to~\cite{Vasak:2023tiu}, the parameters are related via $\ell=1/3M$. For a more in-depth discussion see \cite{paulitype}.}. 
 This quadratic Lagrangian has first been introduced by Gasiorowicz in \cite[p. 90]{gasiorowicz66} in the context of electromagnetism. He added the surface term $i\ell \partial_\mu \bar{\psi} \sigma^{\mu\nu} \partial_\nu \psi$ to the linear Dirac Lagrangian that does not modify the field equation of the free fermion. After gauging the theory, though, an additional anomalous interaction term arises with the coupling constant $\ell$. In the case of electromagnetism, this is the Pauli coupling $e\ell F_{\alpha\beta}\sigma^{\alpha\beta}\psi$, which can be used to restrict the parameter for the electron to $\ell_{e^-}<10^{-8}\, \text{fm}$~\cite{pauliQED}. Gasiorowicz uses this to show, that deducing the Lagrangian from the field equation is ambiguous and has to be treated with care.
Here, the quadratic term is a necessary part of the Dirac Lagrangian, which is otherwise degenerate and does not posses a Hamiltonian representation.
For the gauged Gasiorowicz Lagrangian, the Dirac equation turns out to be \cite[eq. 123f]{Vasak:2023tiu}:
\begin{align}\label{dirac}
i&\Big(\gamma^\beta -\ell\sigma^{\xi\alpha}S^\beta{}_{\xi\alpha}\Big)\Dderr_\beta \psi-\left(m+\frac{\ell}{8}\sigma^{\alpha\beta}\sigma^{n m}R_{nm\alpha\beta}\right)\psi=0.
\end{align}
with the covariant spinor derivative $\Dderr_\mu := \partial_\mu +\frac{i}{4}\sigma^i{}_j\omega^j{}_{i\mu}$, and the spacetime-dependent Dirac  matrices that are built from the standard Dirac matrices by multiplication with the vierbeins. 
 $S^j{}_{\mu\nu}:= e^j{}_\alpha \Gamma^\alpha{}_{[\mu\nu]}$ is the torsion tensor~\eqref{def:torsion}.
 
\medskip
Now the variation of the action integral~\eqref{def:actionintegral0} with respect to the momentum fields gives the canonical equations
\begin{align}
R^i{}_{j\nu\mu}=2\diff{\tilde{\mathcal{H}}_\text{Gr}}{\tilde{q}_i{}^{j\mu\nu}},\qquad S^i{}_{\mu\nu}=\diff{\tilde{\mathcal{H}}_\text{Gr}}{\tilde{k}_i{}^{\mu\nu}},
\end{align}
relating the canonical momenta to the field strengths: 
\begin{subequations}
\begin{align}
\tilde{q}^j{}_{i\mu\nu}&=\varepsilon g_1 (R^j{}_{i\mu\nu}-\hat{R}^j{}_{i\mu\nu}),\\
\tilde{k}^j{}_{\mu\nu}&=\varepsilon g_3 S^j{}_{\mu\nu}.
\end{align}
\end{subequations}
$\hat{R}^j{}_{i\mu\nu}=g_2(e^j{}_\mu g_{\nu\lambda}-e^j{}_\nu g_{\mu\lambda})e_i{}^\lambda$ is the Riemann tensor of maximally symmetric (de~Sitter) spacetime.\par
The field dynamics is derived by combining the canonical equations and considering their symmetric and anti-symmetric portions. 
Thereby the torsion tensor turns out to be totally anti-symmetric. It is found \cite[eq. 134f, 141]{Vasak:2023tiu}
\begin{subequations}
\begin{gather}
S^{\alpha\beta\mu}=S^{[\alpha\beta\mu]},\\
S^\alpha{}_{\nu\mu;\alpha}=-\frac{1}{2g_1g_2+g_3}T_{\text{D}[\nu\mu]}.\label{tordyn}
\end{gather} 
\end{subequations}
The term on the right-hand side of equation (\ref{tordyn}) is the anti-symmetric part of the fermion stress-energy tensor
\begin{equation}\label{def:emD}
 T_\text{D}^{\nu\mu} := -g^{\nu\alpha} e^i{}_\alpha\pfrac{\LCd_{DG}}{e^i{}_\mu},
\end{equation}
which evaluates to~\cite[eq. 110]{Vasak:2023tiu}:
\begin{align}
T_{\text{D}[\nu\beta]}&=\frac{i}{4}\Big(\bar{\psi}\gamma_\beta\Dderr_\nu\psi-\bar{\psi}\gamma_\nu \Dderr_\beta \psi-\bar{\psi}\Dderl_\nu \gamma_\beta\psi+\bar{\psi}\Dderl_\beta\gamma_\nu\psi\Big)\nonumber\\
&\quad+\frac{i\ell}{2}\bar{\psi}\Dderl_\alpha\Big(\sigma_\beta{}^\lambda\delta_\nu^\alpha-\sigma_\nu{}^\lambda\delta_\beta^\alpha-\sigma_\beta{}^\alpha\delta_\nu^\lambda+\sigma_\nu{}^\alpha\delta_\beta^\lambda\Big)\Dderr_\lambda \psi.\label{anti-symmEM}
\end{align}
At first, one may think, that equation (\ref{tordyn}) suffices to derive the dynamics of torsion. However, upon expressing torsion in terms of an axial vector $S^{\alpha\beta\gamma}=\varepsilon^{\alpha\beta\gamma\delta} s_\delta /6$, the left hand side reduces to $\frac{1}{3!}\nabla_\alpha \varepsilon^{\alpha\mu\nu\kappa} s_\kappa = \frac{1}{12}\varepsilon^{\mu\nu\alpha\kappa}(\partial_\alpha s_\kappa-\partial_\kappa s_\alpha)=\frac{1}{12}\varepsilon^{\mu\nu\alpha\kappa} \dt s (\partial_\alpha,\partial_\kappa)$, where $\dt$ is the exterior derivative. In a vacuum, this means, that $s$ is a closed 1-form and thus exact, so it can be written as $s_\mu=\partial_\mu\Phi$. The matter side gives an additional non-conservative contribution, which is connected to the angular momentum of the fermions. The key takeaway however is, that this does not provide a sufficient dynamical description of torsion but is merely an additional constraint.\par
For the dynamics of the curvature tensor, we get \cite[eq. 146]{Vasak:2023tiu}:
\begin{align}\label{curdyn}
-g_1&\left(R^{\nu\beta\mu\alpha}{}_{;\alpha}-R^{\nu\beta\xi\alpha} S^\mu{}_{\xi\alpha}\right)+\left(2g_1g_2+g_3\right)S^{\nu\beta\mu}=-\Sigma^{\nu\beta\mu}.
\end{align}
The term on the right-hand side is the spin tensor \cite[eq. F.4]{Vasak:2023tiu}:
\begin{align}
\Sigma^{ij\beta}:=
\eta^{ik}\,\pfrac{\LCd_{DG}}{\omega^k{}_{j\mu}}
=
&\frac{1}{8}\bar{\psi}\left(\sigma^{ij}\gamma^\beta+\gamma^\beta \sigma^{ij}\right)\psi-\frac{\ell}{4}\bar{\psi}\left(\sigma^{ij}\sigma^{\alpha\beta}\Dderr_\alpha-\Dderl_\alpha \sigma^{\alpha\beta}\sigma^{ij}\right)\psi.\label{spintensor}
\end{align}
Finally, we have the so called CCGG equation, extending Einstein's field equations by quadratic curvature and torsion concomitants \cite[eq. 137]{Vasak:2023tiu}: 
\begin{align}\label{CCGG}
-T_\text{G}^{(\mu\nu)} := 
g_1 &\left(R_{\alpha\beta\gamma}{}^\mu R^{\alpha\beta\gamma\nu}-\tfrac{1}{4} g^{\mu\nu} R_{\alpha\beta\gamma\delta}R^{\alpha\beta\gamma\delta}\right)\nonumber\\
&+\frac{1}{8\pi G}\left(R^{(\mu\nu)}-\tfrac{1}{2}g^{\mu\nu}R -\lambda_0 g^{\mu\nu}\right)\nonumber \\
&-g_3\left(S_{\alpha\beta}{}^\mu S^{\alpha\beta\nu} -\tfrac{1}{2}g^{\mu\nu}S_{\alpha\beta\gamma}S^{\alpha\beta\gamma}\right)=
T_\text{D}^{(\mu\nu)}.
\end{align}
The free parameters $g_i$ in \eqref{def:HGR} have to satisfy $2g_1g_2=1/8\pi G=M_\text{P}^2$ to comply with standard gravitation theory in the weak field limit. 
The constant $\lambda_0=3g_2=3M_\text{P}^2/2g_1$ is a geometrical contribution to the cosmological constant in addition to the vacuum contributions of matter and gravity, and torsion facilitating a dark energy term in the form of a running cosmological ``constant''~\cite{Vasak:2022gps}. 
$T_\text{D}^{(\mu\nu)}$ is the symmetrized stress-energy tensor~\eqref{def:emD} of the Dirac field.
The l.h.s. of the field equation~(\ref{CCGG}) is derived as the negative energy-momentum tensor of gravity,
\begin{equation}\label{def:emG}
 T_\text{G}^{\nu\mu} := -g^{\nu\alpha} e^i{}_\alpha\pfrac{\LCd_{G}}{e^i{}_\mu},
\end{equation}
giving formally a zero-energy-momentum condition for the Universe~\cite{Vasak:2022gps}
\begin{equation}\label{eq:zero-energy-universe}
    T_\text{G}^{(\mu\nu)} + T_\text{D}^{(\mu\nu)} =0.
\end{equation}
Moreover, equations (\ref{curdyn}) and (\ref{CCGG}) 
show that since in this formulation the action is necessarily quadratic in the curvature tensor, 
spacetime responds to matter dynamically with its own inertia.\par

\section{Torsion waves}
In order to obtain a wave equation for the torsion, we start with equation (\ref{curdyn})
and decompose the general connection into the sum of the Levi-Civita connection and the contortion tensor.
In the present case of a totally anti-symmetric torsion, contortion coincides with the torsion tensor:
\begin{align}
\Gamma^\alpha{}_{\beta\gamma}=\bar{\Gamma}^\alpha{}_{\beta\gamma}+S^\alpha{}_{\beta\gamma}.
\end{align}
(From now on, all entities with an overbar are defined with respect to the Levi-Civita connection and do not include torsion.)  
Since we wish to analyse wave-like behaviour, we assume that all components of the torsion tensor are small compared to its change, $S^{\beta\gamma\mu} S_{\mu\alpha}{}^\delta \ll \nabla_\alpha S^{\beta\gamma\delta} $.
Then all terms of quadratic and higher order can be neglected, and the Riemann-Cartan tensor is approximated by
\begin{align}\label{eq:RCapprox}
R^{\nu\beta\mu\alpha}=\bar{R}^{\nu\beta\mu\alpha}+\bar{\nabla}^\mu S^{\nu\beta\alpha}-\bar{\nabla}^\alpha S^{\nu\beta\mu} +\mathcal{O}(S^2),
\end{align}
where $\bar{R}^{\nu\beta\mu\alpha}$ is the torsion-free Riemann curvature tensor.\par
Then, in this limit, taking the divergence of equation~(\ref{eq:RCapprox}) gives
\begin{align}
R^{\nu\beta\mu\alpha}{}_{;\alpha}&=\phantom{+}\bar{\nabla}_\alpha \bar{R}^{\nu\beta\mu\alpha}+g^{\mu\gamma}\bar{\nabla}_\alpha\bar{\nabla}_\gamma S^{\nu\beta\alpha}-\bar{\nabla}_\alpha \bar{\nabla}^\alpha S^{\nu\beta\mu}\nonumber\\
&\quad+ S^\nu{}_{\rho\alpha}\bar{R}^{\rho\beta\mu\alpha}+S^\beta{}_{\rho\alpha} \bar{R}^{\nu\rho\mu\alpha}+S^\mu{}_{\rho\alpha}\bar{R}^{\nu\beta\rho\alpha}+S^\alpha{}_{\rho\alpha}\bar{R}^{\nu\beta\mu\rho}.
\end{align}
Due to anti-symmetry of the torsion tensor, the last term vanishes. For the second term, we use the identity $[\bar{\nabla}_\alpha,\bar{\nabla}_\beta] Z^\gamma=\bar{R}^\gamma{}_{\delta\alpha\beta} Z^\delta$, and its generalization to higher rank tensors. 
With this, we obtain in first order in $S$
\begin{align}
R^{\nu\beta\mu\alpha}{}_{;\alpha}&=
\phantom{+}\bar{\nabla}_\alpha \bar{R}^{\nu\beta\mu\alpha}+g^{\mu\gamma} \bar{\nabla}_\gamma \bar{\nabla}_\alpha S^{\nu\beta\alpha}-
g^{\alpha\gamma} \bar{\nabla}_\alpha \bar{\nabla}_\gamma S^{\nu\beta\mu}\nonumber\\
&\quad+ g^{\mu\gamma}\left(\bar{R}^\nu{}_{\rho\alpha\gamma}S^{\rho\beta\alpha}
-\bar{R}^\beta{}_{\rho\alpha\gamma} S^{\rho\nu\alpha}
+\bar{R}^\alpha{}_{\rho\alpha\gamma}S^{\nu\beta\rho}\right)\nonumber\\
&\quad+ S^\nu{}_{\rho\alpha}\bar{R}^{\rho\beta\mu\alpha}
-S^\beta{}_{\rho\alpha}\bar{R}^{\rho\nu\mu\alpha}
+S^\mu{}_{\rho\alpha}\bar{R}^{\nu\beta\rho\alpha}\nonumber\\
&=\phantom{+}\bar{\nabla}_\alpha \bar{R}^{\nu\beta\mu\alpha}-\frac{1}{M_\text{P}^2+g_3}\bar{\nabla}^\mu T_\text{D}^{[\nu\beta]}-\bar{\nabla}_\alpha\bar{\nabla}^\alpha S^{\nu\beta\mu}\nonumber\\
&\quad+2 \bar{R}^{\rho\beta\mu\alpha}S^\nu{}_{\rho\alpha}
-2\bar{R}^{\rho\nu\mu\alpha}S^\beta{}_{\rho\alpha}
+\bar{R}^{\nu\beta\rho\alpha}S^\mu{}_{\rho\alpha}+\bar{R}_\rho{}^\mu S^{\nu\beta\rho}.
\end{align}
Notice the anti-symmetry in $\nu\beta$. 
At the second equal sign, equation (\ref{tordyn}) was inserted. 
Since the new term involving the energy momentum tensor does not contain any dynamic torsion dependence, it is to be interpreted as a source and we will shift it to the right-hand side to deal with it later.\\
The left-hand side of equation (\ref{curdyn}) thus becomes (after dividing by $g_1$) 
\begin{align}
&-\bar{\nabla}_\alpha \bar{\nabla}^\alpha S^{\nu\beta\mu} 
+2 \bar{R}^{\rho\beta\mu\alpha}S^\nu{}_{\rho\alpha}
-2\bar{R}^{\rho\nu\mu\alpha}S^\beta{}_{\rho\alpha}\nonumber\\
&\quad+\bar{R}_\rho{}^\mu S^{\nu\beta\rho}+\bar{\nabla}_\alpha \bar{R}^{\nu\beta\mu\alpha}-\frac{M_\text{P}^2+g_3}{g_1}S^{\nu\beta\mu}.
\end{align}
\indent To proceed further, we explicitly use the anti-symmetry of the torsion by expressing it in form of an equivalent axial (co-)vector via 
\begin{equation*}
S^{\alpha\beta\gamma}=:\varepsilon^{\alpha\beta\gamma\delta}\frac{s_\delta}{3!}.
\end{equation*}
Furthermore, all three free indices can be contracted with another Levi-Civita tensor and  the fact utilized that, due to metric compatibility, the covariant derivative of the Levi-Civita tensor vanishes. 
In particular, we use
\begin{allowdisplaybreaks}
\begin{subequations}
\begin{align}
\varepsilon_{\xi\nu\beta\mu} S^{\nu\beta\mu}&=s_\xi,\\
\varepsilon_{\xi\nu\beta\mu}\bar{R}^\nu{}_{\rho\alpha}{}^\mu \varepsilon^{\rho\beta\alpha\kappa}s_\kappa&=\bar{R}^\nu{}_{\rho\alpha}{}^\mu s_\kappa\big(\delta^\rho_\xi \delta^\alpha_\nu \delta^\kappa_\mu +\delta^\kappa_\xi \delta^\rho_\nu \delta^\alpha_\mu +\delta^\alpha_\xi \delta^\kappa_\nu \delta^\rho_\mu\nonumber\\*
&\qquad-\delta^\alpha_\xi\delta^\rho_\nu \delta^\kappa_\mu -\delta^\kappa_\xi \delta^\alpha_\nu\delta^\rho_\mu -\delta^\rho_\xi\delta^\kappa_\nu\delta^\alpha_\mu\big)/6\nonumber\\*
&= \left(\bar{R}^\alpha{}_{\xi\alpha}{}^\mu s_\mu +\bar{R}^\kappa{}_{\rho\xi}{}^\rho s_\kappa -\bar{R}^\alpha{}_{\mu\alpha}{}^\mu s_\xi\right)/6\nonumber\\*
&=\left(2\bar{R}_\xi{}^\mu s_\mu -\bar{R} s_\xi\right)/6,\\
\varepsilon_{\xi\nu\beta\mu}\bar{R}_\rho{}^\mu \varepsilon^{\nu\beta\rho\kappa}s_\kappa&=-2\bar{R}_\rho{}^\mu s_\kappa \big(\delta^\rho_\xi \delta^\kappa_\mu -\delta^\kappa_\xi \delta^\rho_\mu\big)/6\nonumber\\*
&=\left(2\bar{R}s_\xi-2\bar{R}_\xi{}^\mu s_\mu\right)/6,\\
\varepsilon_{\xi\nu\beta\mu}\bar{R}^{\nu\beta\mu\alpha}&=0.
\end{align}
\end{subequations}
\end{allowdisplaybreaks}
For the first three equations, we used the well known identities for the contraction of two Levi-Civita tensors \cite[1.1.30]{poplawski2020classical}, while the last one results from the Bianchi identity. 
After dividing by $g_1$, we obtain the following wave equation: 
\begin{align}
-\bar{\nabla}_\alpha\bar{\nabla}^\alpha &s_\xi+\bar{R}_\xi{}^\mu s_\mu-\left(\frac{\bar{R}}{3}+\frac{M_\text{P}^2+g_3}{g_1}\right)s_\xi=\frac{1}{g_1}\varepsilon_{\xi\nu\beta\mu}\Sigma^{\nu\beta\mu}-\frac{1}{g_1(M_\text{P}^2+g_3)}\varepsilon_{\xi\nu\beta\mu}\bar{\nabla}^\mu T_\text{D}^{[\nu\beta]}.
\end{align}
\indent The first two terms on the left-hand side can be combined into the so-called \mbox{deRham} Laplacian $\bar{\Delta}_\text{(dR)} v_\alpha=\bar{\nabla}_\beta \bar{\nabla}^\beta v_\alpha-\bar{R}_\alpha{}^\beta v_\beta$, which is the proper generalization of the Laplace operator acting on d-forms in curved spacetimes \cite{deRhamII,deRhamI}.\\
Then the final form of the wave equation for a massive axial torsion vector field sourced by fermion spin becomes: 
\begin{align}\label{waveEq}
\bar{\Delta}_{(\text{dR})}&s_\xi+\left(\frac{M_\text{P}^2+g_3}{g_1}+\frac{\bar{R}}{3}\right)s_\xi=-\frac{1}{g_1}\varepsilon_{\xi\nu\beta\mu}\Sigma^{\nu\beta\mu}+\frac{1}{g_1(M_\text{P}^2+g_3)}\varepsilon_{\xi\nu\beta\mu}\bar{\nabla}^\mu T_\text{D}^{[\nu\beta]}.
\end{align}
\section{Sources of torsion waves}
In order to understand the nature of the sources of torsion we analyze the right-hand side of equation (\ref{waveEq}). 
In the weak-torsion limit we assume no back-reaction of the torsion, hence all torsion related terms are neglected here. To keep good readability, we do not explicitly write out the over-bar on the spinor covariant derivative. \par
The spin tensor is then approximated by
\begin{align}
\Sigma^{\nu\beta\mu}&=-\frac{1}{4}\varepsilon^{\nu\beta\mu\kappa}\bar{\psi}\gamma^5\gamma_\kappa  \psi-\frac{\ell}{4}\left( \bar{\psi}\sigma^{\nu\beta}\sigma^{\mu\kappa}\Dderr_\kappa \psi+\bar{\psi}\Dderl_\kappa \sigma^{\mu\kappa} \sigma^{\nu\beta}\psi\right),
\end{align}
where we have used the identity $\gamma^a\sigma^{cb}+\sigma^{cb}\gamma^a=-2\varepsilon^{abcd}\gamma_d\gamma^5$ \cite[eq. 4.201]{geomAlg}.\par
Since the parameter $\ell$ is restricted to be very small by particle physics related experiments \cite{pauliQED}, we will neglect all contributions proportional to it at this point. For the sake of completeness, the full computation can be found in Appendix \ref{appGasi}.\par
Performing the contraction with the Levi-Civita tensor, we obtain
\begin{align}
\varepsilon_{\xi\nu\beta\mu}\Sigma^{\nu\beta\mu}&=-\frac{3}{2}\bar{\psi} \gamma^5\gamma_\xi \psi
\end{align}
\indent The second contribution is the derivative of the anti-symmetric part of the stress-energy  tensor. Using the Leibniz rule and once again neglecting terms of $\mathcal{O}(\ell)$, it is given by
\begin{align}
\bar{\nabla}_\mu T_{\text{D}[\nu\beta]}&=\phantom{+}\frac{i}{4}\bar{\psi}\Big(\Dderl_\mu\gamma_\beta\Dderr_\nu-\gamma_\beta \bar{\Gamma}^\xi{}_{\nu\mu}\Dderr_\xi+\gamma_\beta\Dderr_\mu\Dderr_\nu \Big)\psi\nonumber\\
&\phantom{=}-\frac{i}{4}\bar{\psi}\Big(\Dderl_\mu\gamma_\nu\Dderr_\beta-\gamma_\nu\bar{\Gamma}^\xi{}_{\beta\mu}\Dderr_\xi+\gamma_\nu\Dderr_\mu\Dderr_\beta\Big)\psi\nonumber\\
&\phantom{=} +\frac{i}{4}\bar{\psi}\Big(\Dderl_\beta\Dderl_\mu \gamma_\nu -\Dderl_\xi \bar{\Gamma}^\xi{}_{\beta\mu}\gamma_\nu +\Dderl_\beta\gamma_\nu \Dderr_\mu\Big)\psi\nonumber\\
&\phantom{=}-\frac{i}{4}\bar{\psi}\Big(\Dderl_\nu\Dderl_\mu \gamma_\beta -\Dderl_\xi \bar{\Gamma}^\xi{}_{\nu\mu}\gamma_\beta +\Dderl_\nu\gamma_\beta \Dderr_\mu\Big)\psi.
\end{align}
An exemplary computation for the first term can be found in Appendix \ref{appTD}. When contracting with a Levi-Civita tensor, the terms explicitly involving the Christoffel symbols vanish due to their symmetry in the lower index pair. Rearranging the remaining terms gives then 
\begin{align}
\varepsilon_\xi{}^{\mu\nu\beta}\bar{\nabla}_\mu T_{\text{D}[\nu\beta]}&= \varepsilon_\xi{}^{\mu\nu\beta}\frac{i}{2}\bar{\psi}\Big(\Dderl_\beta \Dderl_\mu \gamma_\nu +\gamma_\beta \Dderr_\mu\Dderr_\nu \Big)\psi\nonumber\\
&\phantom{=}+i\varepsilon_\xi{}^{\mu\nu\beta}\bar{\psi}\Dderl_\beta\gamma_\nu\Dderr_\mu \psi
\end{align}
To simplify the first line, we can introduce the Clifford algebra valued Riemann tensor defined by the commutator of the spin covariant derivative:
\begin{align}
\left[\Dderr_\beta,\Dderr_\mu\right]=\left[\Dderl_\mu,\Dderl_\beta\right]=\frac{i}{4}R^i{}_{j\beta\mu}\sigma^j{}_i=:\mathbf{R}_{\beta\mu}.
\end{align}
(The derivation can be found in Appendix \ref{appRep}.) Contracting with the Levi-Civita tensor gives
\begin{align}
\varepsilon_\xi{}&^{\nu\beta\mu}\bar{\nabla}_\mu T_{\text{D}[\nu\beta]}=\frac{i}{4}\varepsilon_\xi{}^{\nu\beta\mu}\bar{\psi}\left(\bar{\mathbf{R}}_{\mu\beta}\gamma_\nu-\gamma_\nu \bar{\mathbf{R}}_{\mu\beta}+4\Dderl_\beta\gamma_\nu\Dderr_\mu \right)\psi
\end{align}
For the commutator of the Riemann tensor and the gamma matrix we have
\begin{align}
[\bar{\mathbf{R}}_{\mu\beta},\gamma_k]e^k{}_\nu&=\frac{i}{4}\bar{R}^i{}_{j\mu\beta} e^k{}_\nu [\sigma^j{}_i,\gamma_k]\nonumber\\
&=\frac{i}{4}\bar{R}^i{}_{j\mu\beta} e^k{}_\nu 2i(\eta_{ik}\gamma^j-\delta^j_k\gamma_i)\nonumber\\
&=-\frac{1}{2}\left(\bar{R}_{\nu j\mu\beta}\gamma^j-\bar{R}^i{}_{\nu\mu\beta}\gamma_i\right)=\gamma_i \bar{R}^i{}_{\nu\mu\beta}.
\end{align}
Because of the first Bianchi identity, this term vanishes upon contraction with the permutation tensor. \par
The entire source of torsion waves is thus given by
\begin{align}
Q_\xi&=\frac{3}{2g_1} \bar{\psi} \gamma^5\gamma_\xi\psi-\frac{i}{g_1(M_\text{P}^2 +g_3)}\varepsilon_\xi{}^{\beta\nu\mu}\bar{\psi}\Dderl_\beta\gamma_\nu\Dderr_\mu\psi
\end{align}
\indent In the second term, the numerator is of the order of the energy of the particle squared, while the denominator is $\mathcal{O}(M_\text{P}^2)=(10^{19}\text{GeV})^2$, so that this term may safely be neglected.\par
The only term remaining is the first one,  the axial current of the Dirac field.
\begin{align}
Q_\xi=\frac{3}{2g_1}\bar{\psi}\gamma^5\gamma_\xi\psi=\frac{3}{2g_1} \,j^\text{A}_\xi.
\end{align}
The axial current denotes the net spin of a system, which is effectively the difference between spin-up and spin-down states.\par
The entire wave equation results as 
\begin{align}\label{waveeqn}
\bar{\Delta}_{(\text{dR})}s_\xi+\left(\frac{\bar{R}}{3}+\frac{M_\text{P}^2+g_3}{g_1}\right)s_\xi=\frac{3}{2g_1}j_\xi^\text{A}.
\end{align}
Before we come to the discussion, we should take a closer look at the effective mass term. By taking the trace of the CCGG equation (\ref{CCGG}), we obtain
\begin{align}
\bar{R}=-8\pi G(T_\text{M}+ 6 (g_3+M_\text{P}^2) s_\xi s^\xi) -4\lambda_0.
\end{align}
$T_\text{M}$ is the trace of the energy-momentum tensor of matter and includes the vacuum energy contribution,~$T_\text{M} = T_\text{vac}+\tilde{T}_\text{M}$.
Now due to the zero-energy condition \eref{eq:zero-energy-universe}, the vacuum energy of matter is canceled by the vacuum energy of spacetime, giving $T_\text{vac}=-3M_\text{P}^4/(2 g_1)$, see~\cite{Vasak:2022gps}.
Then
\begin{align}
\bar{\Delta}_{(\text{dR})}s_\xi+\left(\frac{g_3+M_\text{P}^2}{g_1}-\frac{1}{3M_\text{P}^2}T_\text{M}\right)s_\xi=\frac{3}{2g_1}j_\xi^\text{A},
\end{align}
and the torsion waves emerge with the effective mass
\begin{align}
    m^2=\frac{M_\text{P}^2+g_3}{g_1}-\frac{1}{3 M_\text{P}^2}\tilde{T}_\text{M}.
\end{align}
$\tilde{T}_\text{M}$ is the trace of the normal-ordered energy momentum tensor of matter, i.e.~without vacuum energy contributions. In cosmology the trace of the energy-momentum tensor is just the average (dark) matter density, $T_\text{M} = \rho_\text{m}$, and the second term thus gives a small contribution~$\frac{1}{3 M_\text{P}^2}\rho_\text{m}$.  

\section{Discussion}\label{discu}
The parameter $g_1$ determines the mass as well as the strength of the excitation of the torsion field. 
The value of this constant remains uncertain, though, since not much research has been put into restricting the constants of the theory yet.  
In cosmological studies~\cite{cosmology}, $g_1$ has been estimated to be of the order $10^{120}$ yielding extremely weak excitations.
That estimate, however, has to be taken with a grain of salt since the parameters used in that paper are mainly educated guesses, and a thorough analysis might require their substantial adjustment.\par
\medskip
Demanding a real mass of the torsion field even in neutron stars, where $\tilde{T}_\text{M}\approx 10^{17} \text{kg}/\text{m}^3$, we obtain, when assuming $g_3\lesssim M_\text{P}^2$, the upper bound $g_1< M_\text{P}^4/\tilde{T}_\text{M}\approx 10^{80}$. 
Notice that for the naive estimate of the vacuum energy, $T_\text{vac} \sim M^4_\text{P}$ as suggested in~\cite{weinberg89}, 
$g_1$ is of the order of unity and negative,
leading to tachyonic torsion waves and implying an instability of the system in the low torsion regime. 
Nevertheless, in that case the torsion field would obtain a quartic potential similar to the Higgs field, which would have an unstable extremum at $s=0$ and a minimum at a non-zero field value. 
This value can also be obtained from the curvature dynamics equation~\eqref{curdyn} upon neglecting all interactions of the torsion field (including that with the curved background), and assuming constant torsion. 
With these assumptions, that curvature equation becomes, after contracting with the Levi-Civita symbol,
\begin{equation}
    0=\left(\frac{s^2}{18}+\frac{M_\text{P}^2+g_3}{g_1}\right)s_\gamma 
\end{equation}
with the solution $s^2=-18(M_\text{P}^2+g_3)/g_1$. This is of course only relevant in the case $g_1<0$ since otherwise, there is no minimum in the potential.
This leads, just like in the Higgs case, to spontaneous symmetry breaking via a non-vanishing vacuum expectation value~$s^\nu_\text{vac} \equiv v^\nu$ of torsion. Such a solution raises many questions, though. On the one hand, in general curved space times, there does not exist a constant vector field with $\bar{\nabla}_\mu v^\nu=0$, so even the existence of such a vacuum expectation value in curved spaces is unclear. Additionally, such a solution would necessarily break Lorentz invariance due to its orientation, which also raises many questions including the existence of Goldstone bosons. All this will be examined in a future paper.\par
\medskip
A crucial aspect for generating and detecting torsion 
is the interaction of torsion and spin. 
It remains an open question for further research, how to measure torsion waves and what might be a mechanism to create a net spin current that can generate torsion waves with detectable intensity somewhere in the universe (or even in a lab experiment). \par 
A possible source of torsion waves could be neutron stars  exhibiting spin polarised states in their interior \cite{neutronPolIII}\cite{neutronPolII}\cite{neutronPolI}. Two such colliding neutron stars with  opposing polarisation states would cause oscillations of the axial current, and thus generate potentially observable torsion waves. 
On the other hand, even with the polarization directions of the neutron stars aligned, quadrupole radiation would be emitted, albeit significantly weaker. 
Inspiraling patterns of neutron star mergers could allow to measure the influence of torsion waves indirectly by calculating the impact of the energy radiated away by torsion waves.
Torsion waves could also be generated by the strong magnetic fields of magnetars causing matter spin to align and strong net spin currents to emerge.\par
\medskip
Before being able to set up experiments for measuring torsion directly, we have to understand how torsion waves or torsion in general influence the behaviour of matter. 
This requires, in the first place, an in-depth analysis of the modified Dirac equation, e.g. by using the WKB approximation~\cite{WKB} or a transport theory approach~\cite{vasak87,transportTheory}.\par
In order to detect torsion waves, one could for example look at differences in the trajectories of neutrinos and photons emitted by pulsars. 
 Since fermions are affected by torsion but bosons are not, this might lead to deviations in arrival times or to apparently different locations of the source due to the deflection by the torsion waves. 
 Additionally, an experiment similar to the pulsar timing array might be feasible. 
 Focusing on the neutrino radiation of pulsars, an analogous experiment might 
 lead to the detection of a torsion wave background. 
 And a further possible novel effect of torsion might be the interaction of neutrinos with the background radiation causing anomalous flavour oscillations~\cite{neutrinomixing}.

\section{Conclusion}

In the course of this paper, we have shown that in the weak-field limit torsion may exhibit wave-like behaviour upon small excitations which are caused by fermionic axial currents, the net spin density 4-current of a system. This aligns with the intuition that spin causes a ``twisting'' of spacetime.\par
Based on this, numerous possible mechanisms for generation and detection of torsion waves were discussed, potentially facilitating a new field of torsion-based astronomy. 

\section*{Acknowledgements}
The authors are indebted to the Walther-Greiner-Gesellschaft f\"{u}r physikalische Grundlagenforschung, and especially to the Fueck Foundation in Frankfurt for their support. They also wish to thank the unknown referee for their constructive comments and helpful suggestions, and  Laura Sagunski, Armin van den Venn  and Keiwan Jamaly for insightful discussions.

\bibliography{bib}{}
\thispagestyle{empty}

\appendix

\section{Canonical transformation theory in De Donder Weyl Hamiltonian formulation}\label{appDWHtrafo}

In this formulation~\cite{struckmeier08, Vasak:2023iaz, Vasak:2023ncz, Vasak:2023tiu} all four spacetime dimensions are considered on equal footing, in contrast to standard field theory where the time direction is singled out. 
Formally the existence of a non-degenerate Lagrangian density $\tilde{\LCd} := \sqrt{-g}\,{\LCd}$ is assumed such that the so called De Donder-Weyl (DW) Hamiltonian can be constructed via a Legendre transformation involving a covariant set of "momentum" fields. The factor $\sqrt{-g}$  transforms the Lagrangian scalar into a scalar density and converts the action into a world scalar.
For a real scalar field $\phi$ this means, for example, that the canonical momentum field is defined as
\begin{equation*}
 \tilde{\pi}^\mu := \pfrac{\tilde{\LCd}(\phi,\phi_{,\mu})}{\phi_{,\mu}}.
\end{equation*}
Comma denotes the partial derivative with respect to $x$. 
Now Legendre transforming the Lagrangian density gives the De Donder-Weyl (DW) Hamiltonian
\begin{equation}\label{def:Hsimple}
\tilde{\HCd}(\phi,\tilde{\pi}^\mu) := \tilde{\pi}^\mu\,\phi_{,\mu} - \tilde{\LCd},
\end{equation}
and the action integral becomes
\begin{equation}
S =  \int_{V}\tilde{\LCd}\,\dd^4x
=
\int_{V}\left(\tilde{\pi}^\mu\,\phi_{,\mu} - \tilde{\HCd}\right)\,\dd^4x.
\end{equation}
The variation of the action integral w.r.t. the now independent conjugate fields $\phi$ and $\tilde{\pi}^\mu$ leads to the canonical equations 
\begin{subequations}
 \begin{align}\label{eq:canonicalequations0}
\phi_{,\nu}&=\pfrac{\tilde{\HCd}}{\tilde{\pi}^\nu} \\
\tilde{\pi}\indices{^\nu_{,\nu}}&=-\pfrac{\tilde{\HCd}}{\phi}.
\end{align}
\end{subequations}

Curved spacetimes are described as principal bundles in differential geometry, a manifold $M$. 
While elements of that manifold, points or events, are considered physical entities, their coordinates are mere labels that can be arbitrarily chosen. 
That arbitrariness, coined Principle of General Relativity by Einstein, corresponds to invariance of any physical theory with respect to arbitrary (passive) diffeomorphisms.
Matter fields are sections on the tangent space of that bundle, and the geometry of spacetime is represented by the vierbein (co-vector) fields $e\indices{^i_\mu}$ that determine the metric via
\begin{equation}
g_{\mu\nu} \equiv \eta_{ij}\, e\indices{^i_\mu}\,e\indices{^j_\nu}.
\end{equation}
Vierbeins build a basis of an inertial spacetime equipped with the Minkowski metric $\eta_{ij} =$ diag(1,-1,-1,-1) that is attached at each point of the bundle. Those frames are elements of a "fiber" w.r.t. (orthochonous) Lorentz transformations $\Lambda\indices{^I_i}(x) \in$~SO($1,3$).
Here Latin indices relate to the inertial frame, while Greek indices are components in the base manifold, both assuming values in $\{0,1,2,3\}$. 
For a scalar field embedded in curved spacetime the corresponding dynamical system $(\varphi(x), e\indices{^i_\nu}(x))$ is thus subject to a (gauge) ambiguity with respect to transformations that are elements of the symmetry group SO($1,3$)$\times$Diff($M$).  
Such a transformation from the original frame, denoted by lower-case letters and indices, to the transformed system, denoted by capital letters and indices, is:
\begin{subequations}
\begin{align}
 \varphi(x) &\mapsto \Phi(X) = \varphi(x) \\
 e\indices{^i_\nu}(x) &\mapsto E\indices{^I_\mu}(X) = \Lambda\indices{^I_i}(x)\,e\indices{^i_\nu}(x)\,\pfrac{x^\nu}{X^\mu}.\label{def:trafotetrad}
\end{align}
\end{subequations}
For the dynamics of the physical system to be "immune" against such an ambiguity the action integral must be invariant up to a boundary term on which the fields are fixed, which for the Lagrangian density means:
\begin{align}
&\tilde{\LCd}^\prime\left(\Phi,\pfrac{\Phi}{X^\nu},E\indices{^I_\mu},\pfrac{E\indices{^I_\mu}}{X^\nu},X\right)\detpartial{X}{x}\\
&\qquad \must\,
\tilde{\LCd}\left(\varphi,\pfrac{\varphi}{x^\nu},e\indices{^i_\mu},\pfrac{e\indices{^i_\mu}}{x^\nu},x\right)-\pfrac{\tilde{\FCd}^\nu}{x^\nu}.\nonumber
\end{align}
Thereby $\tilde{\FCd}^\nu$ is a vector density that facilitates the surface term.
Using the DW Hamiltonian (with the momentum field $\tilde{k}\indices{_i^{\mu\nu}}$ conjugate to vierbein) this gives:
\begin{align}
&\left[\tilde{\Pi}^\nu\,\pfrac{\Phi}{X^\nu} -
\tilde{K}\indices{_I^{\mu\nu}}\,\pfrac{E\indices{^I_\mu}}{X^\nu} -
\tilde{\HCd}^\prime\left(\Phi,\tilde{\Pi}^\nu,E\indices{^I_\mu},\tilde{K}\indices{_I^{\mu\nu}},X\right)\right]
\detpartial{X}{x}\nonumber \\
&\must\,
\tilde{\pi}^\nu\,\pfrac{\phi}{x^\nu} -
\tilde{k}\indices{_i^{\mu\nu}}\,\pfrac{e\indices{^i_\mu}}{x^\nu} -
\tilde{\HCd}\left(\phi,\tilde{\pi}^\nu,e\indices{^i_\mu},\tilde{k}\indices{_i^{\mu\nu}},x\right)
-\pfrac{\tilde{\FCd}^\nu}{x^\nu}.
\end{align}
While the first two terms on both sides of this equation display the appropriate transformation property, the Hamiltonian density must obviously satisfy
the so called canonical transformation rule
\begin{align}\label{eq:Htransform}
&\tilde{\HCd}^\prime\left(\Phi,\tilde{\Pi}^\nu,E\indices{^I_\mu},\tilde{K}\indices{_I^{\mu\nu}},X\right)\detpartial{X}{x} \\
& \qquad  =
\tilde{\HCd}\left(\phi,\tilde{\pi}^\nu,e\indices{^i_\mu},\tilde{k}\indices{_i^{\mu\nu}},x\right)
+\left.\pfrac{\tilde{\FCd}^\nu}{x^\nu}\right|_{\text{expl}}.\nonumber
\end{align}
The vector density $\tilde{\FCd}^\nu$ is the key lever for generating system invariance against transformations of the involved matter fields with respect to a given local symmetry. 
For local SO($1,3$)$\times$Diff($M$) field transformations we set $\tilde{\FCd}^\nu = \tilde{\FCd}_3^\nu$. 
Here $\tilde{\FCd}_3^\nu$ is a generating function that depends on the original momenta and on the transformed fields,
reflecting that the scalar field does not change upon the above symmetry transformation while the vierbein transforms as a vector with respect to both indices:
\begin{equation}
\tilde{\FCd}_3^\nu\left(\Phi,\tilde{\pi}^\nu, E\indices{^I_\mu},
\tilde{k}\indices{_i^{\mu\nu}},x\right)
=-\tilde{\pi}^\nu\,\Phi-\tilde{k}\indices{_i^{\beta\nu}}\,\Lambda\indices{^i_I}\,E\indices{^I_\alpha}\,\pfrac{X^\alpha}{x^\beta}.
\end{equation}
Obviously, the explicit derivative of that generating function in \eref{eq:Htransform} acts on the spacetime-dependent transformation matrices $\pfrac{X^\alpha}{x^\beta}$ and $\Lambda\indices{^i_I}$:
\begin{align}\label{HLorentzandchart}
\left.\pfrac{\FCd_3^\nu}{x^\nu} \right|_{\text{expl}}&=
-\tilde{k}\indices{_i^{\beta\nu}}\pfrac{}{x^\nu}\left(\Lambda\indices{^i_I}\pfrac{X^\alpha}{x^\beta}\right)E\indices{_\alpha^I}\\
&=-\tilde{k}\indices{_i^{(\beta\nu)}}\pfrac{}{x^\nu}\left(\Lambda\indices{^i_I}\pfrac{X^\alpha}{x^\beta}\right)E\indices{_\alpha^I}
-\tilde{k}\indices{_i^{[\beta\nu]}}\pfrac{\Lambda\indices{^i_I}}{x^\nu}\pfrac{X^\alpha}{x^\beta}E\indices{_\alpha^I}.\nonumber
\end{align}
It does not vanish reflecting the lack of the required local symmetry of the original Lagrangian and the corresponding Hamiltonian densities.
Using the partial derivative of the transformation law,~\eref{def:trafotetrad}, the terms $-\tilde{k}\indices{_i^{(\beta\nu)}}\pfrac{e\indices{_\beta^i}}{x^\nu}$ and $\tilde{K}\indices{_I^{(\beta\nu)}} \pfrac{E\indices{_\beta^I}}{X^\nu}\detpartial{X}{x}$ can be combined with similar terms in \eref{eq:Htransform} to give
\begin{align}\label{F3derivative2concr}
&-\pfrac{\tilde{\pi}^\alpha}{x^{\alpha}}\,\varphi
-\pfrac{\tilde{k}\indices{_i^{[\mu\alpha]}}}{x^{\alpha}}\,e\indices{_\mu^i}
-\tilde{\HCd}\left(\varphi,\tilde{\pi}^\nu,e\indices{_\mu^i},\tilde{k}\indices{_i^{\mu\nu}},x\right)\\
&-\left[\tilde{\Pi}^\nu\pfrac{\Phi}{X^{\nu}}+\tilde{K}\indices{_I^{[\mu\nu]}}\pfrac{E\indices{_\mu^I}}{X^\nu}-
\tilde{\HCd}^\prime \left(\Phi,\tilde{\Pi}^\nu,E\indices{_\mu^I},\tilde{K}\indices{_I^{\mu\nu}},X\right)\right]\detpartial{X}{x}\nonumber\\
&\qquad=\tilde{k}\indices{_i^{[\beta\nu]}}\Lambda\indices{^i_I}\pfrac{\Lambda\indices{^I_j}}{x^\nu} e\indices{_\beta^j}.\nonumber
\end{align}
The remaining term on the right-hand side of \eref{F3derivative2concr} contains the spacetime-dependent Lorentz transformation coefficients $\Lambda\indices{^I_j}(x)$.
The only way to re-establish the invariance of the system dynamics is to introduce a "counter term" whose transformation rule absorbs the symmetry-breaking term proportional
to $\partial\Lambda\indices{^I_j}/\partial x^\nu$. 
That new term called gauge Hamiltonian must thus transform as
\begin{equation}\label{eq:inv-cond}
\tilde{\HCd}_{\mathrm{Gau}}^{\prime}\,\detpartial{X}{x}-\tilde{\HCd}_{\mathrm{Gau}}=
\tilde{k}\indices{_i^{[\mu\nu]}}\Lambda\indices{^i_I}\pfrac{\Lambda\indices{^I_j}}{x^\nu}\,e\indices{_\mu^j}.
\end{equation}
The gauge Hamiltonian is chosen such that the {free} indices $i,j,\nu$ of $\Lambda\indices{^i_I}\partial\Lambda\indices{^I_j}/\partial x^\nu$ are exactly matched:
\begin{equation}\label{g-ham1}
\tilde{\HCd}_{\mathrm{Gau}}=-\tilde{k}\indices{_i^{[\mu\nu]}}\,\ho\indices{^i_{j\nu}}\,e\indices{_\mu^j}.
\end{equation}
Thereby the newly introduced gauge field~$\ho\indices{^i_{j\nu}}$ must retain its form when transformed, hence:
\begin{equation}\label{g-ham2}
\tilde{\HCd}_{\mathrm{Gau}}^{\prime}=-\tilde{K}\indices{_I^{[\mu\nu]}}\,\HO\indices{^I_{J\nu}}\,E\indices{^J_\mu}.
\end{equation}
$\HO\indices{^I_{J\nu}}$ is the transformed gauge field, and from the transformation relation~\eqref{eq:inv-cond} it follows that that transformation must be inhomogeneous:
\begin{equation}\label{omegatransform1}
\ho\indices{^i_{j\nu}}=\Lambda\indices{^i_I}\,\HO\indices{^I_{J\alpha}}\,\Lambda\indices{^J_j}\,\pfrac{X^\alpha}{x^\nu}
+\Lambda\indices{^i_I}\,\pfrac{\Lambda\indices{^I_j}}{x^{\nu}}.
\end{equation}
As this is exactly the transformation property of the spin connection, the gauge field can be identified with the spin connection.

The covariant canonical transformation theory thus \emph{derives} gravity as a Yang-Mills type gauge theory wielding
four independent dynamical gravitational fields: the vierbein, $e\indices{^i_\mu}$, representing the geometry, the gauge field spin connection, $\omega\indices{^i_{j\nu}}$, defining parallel transport,
and the respective conjugate momentum fields, $\tilde{k}\indices{_i^{\mu\nu}}$ and $\hoc\indices{_i^{j\mu\nu}}$, defined as:
\begin{equation}
 \tilde{k}\indices{_i^{\mu\nu}} \equiv k\indices{_i^{\mu\nu}}\varepsilon :=
\pfrac{\tilde{\LCd}_{\mathrm{\mathrm{tot}}}}{e\indices{^i_{\mu,\nu}}}   \qquad
\hoc\indices{_i^{j\alpha\beta}} \equiv q\indices{_i^{j\alpha\beta}}\varepsilon
:=
\pfrac{\tilde{\LCd}_{\mathrm{\mathrm{tot}}}}{\omega\indices{^i_{j\alpha,\beta}}}
 \end{equation}
 with $\varepsilon := \det e\indices{^k_\beta} \equiv \sqrt{-\det g_{\mu\nu}}$.

The resulting action integral~\cite[eq. 106]{Vasak:2023tiu} is a world scalar, and the integrand is form-invariant under the transformation group~SO($1,3$)$\times$Diff($M$):
\begin{align}\label{def:actionintegral0}
S_0 &=
\int_{V}\tilde{\LCd}_{\mathrm{\mathrm{tot}}}\,\dd^4x \\
&=
\int_{V}\left(\tilde{k}\indices{_i^{\mu\nu}}\,S\indices{^i_{\mu\nu}}
+\onehalf \,
\hoc\indices{_i^{j\mu\nu}}\,R\indices{^i_{j\mu\nu}}-\tilde{\HCd}_{\mathrm{Gr}}+
\tilde{\LCd}_{\mathrm{\mathrm{matter}}}\right)\dd^4x.\nonumber
\end{align}
Compared to \eref{def:Hsimple}, the field derivatives (``velocities'') of the vierbein and the connection have in the gauging procedure miraculously morphed into covariant field strengths, namely torsion of spacetime and Riemann-Cartan curvature, respectively, defined as:
 \begin{subequations}
\begin{align}
S\indices{^i_{\mu\nu}} :=& \,\onehalf \left(
\pfrac{e\indices{^i_\mu}}{x^\nu} - \pfrac{e\indices{^i_\nu}}{x^\mu}
+ \omega\indices{^i_{j\nu}} \, e\indices{^j_\mu} - \omega\indices{^i_{j\mu}} \,
e\indices{^i_\nu} \right)\\ 
 \equiv& \,e\indices{^i_\lambda}\,S\indices{^\lambda_{\mu\nu}} =  e\indices{^i_\lambda}\,\gamma\indices{^\lambda_{[\mu\nu]}}
\nonumber \label{def:torsion}\\
R\indices{^i_{j\mu\nu}} :=& \,\pfrac{\omega\indices{^i_{j\nu}}}{x^\mu} -
\pfrac{\omega\indices{^i_{j\mu}}}{x^\nu} +
\omega\indices{^i_{n\mu}}\,\omega\indices{^n_{j\nu}} -
\omega\indices{^i_{n\nu}}\,\omega\indices{^n_{j\mu}} \\
\equiv& \, e\indices{^i_\lambda}\,e\indices{_j^\sigma}\,R\indices{^\lambda_{\sigma\mu\nu}} \nonumber \\
=& \,e\indices{^i_\lambda}\,e\indices{_j^\sigma}\,\left(\pfrac{\gamma\indices{^{\lambda}_{\sigma\nu}}}{x^{\mu}} -
 \, \pfrac{\gamma\indices{^{\lambda}_{\sigma\mu}}}{x^{\nu}} +
 \, \gamma\indices{^{\lambda}_{\delta\mu}}\gamma\indices{^{\delta}_{\sigma\nu}} -
 \, \gamma\indices{^{\lambda}_{\delta\nu}}\gamma\indices{^{\delta}_{\sigma\mu}}\right) \nonumber \label{def:curvature}
\,.
\end{align}
\end{subequations}
This identification is achieved as the expression
\begin{equation} \label{def:gammainomega2}
\gamma\indices{^{\mu}_{\alpha\nu}}
:= e\indices{_{k}^{\mu}}
\left(\pfrac{e\indices{_{\alpha}^{k}}}{x^{\nu}} + \omega\indices{^{k}_{i\nu}} \, e\indices{_{\alpha}^{i}}
\right)
\end{equation}
can be identified with the affine connection.
The proof is straightforward since the transformation law for the affine connection,
\begin{equation} \label{gammatransform}
\Gamma\indices{^{\alpha}_{\nu\beta}} \,=\,\gamma\indices{^{\sigma}_{\eta\mu}}\,
  \pfrac{x^\eta}{X^\nu} \,\pfrac{x^\mu}{X^\beta} \,\pfrac{X^\alpha}{x^\sigma}
  \,-\,\pfrac{x^\eta}{X^\nu}\,\pfrac{x^\mu}{X^\beta}\,\ppfrac{X^\alpha}{x^\mu}{x^\eta},
\end{equation}
derives from the transformation law~\eqref{omegatransform1} of the spin connection.
Notice that here and in the following the affine connection coefficients are not independent fields but just a placeholder for the right-hand side\
of the definition \eref{def:gammainomega2}. 

It is useful for a more compact notation to define a covariant derivative on the frame bundle denoted by ``$;$'', that acts on both the Lorentz and coordinate indices.
Then we can re-write the definition~\eqref{def:gammainomega2} as
\begin{equation}\label{def:vierbeinenpostulat3}
e\indices{_\mu^{i}_{;\nu}}=\pfrac{e\indices{_\mu^i}}{x^\nu}+\omega\indices{^i_k_\nu}\,e\indices{_\mu^k}\,-\,
\gamma\indices{^\alpha_\mu_\nu}\,e\indices{_\alpha^i}\equiv0.
\end{equation}
This is called the Vierbein Postulate, which ensures compatibility between objects expressed in the basis of the curved manifold or in the basis of the local inertial frame. Provided the spin connection is anti-symmetric in $ij$, which we shall assume henceforth, this also ensures metric compatibility, i.e.\ the vanishing covariant derivative of the metric and thus the preservation of lenghts and angles,
\begin{equation} \label{def:metricity}
g\indices{_{\mu\nu;\alpha}}(x)=-e\indices{_\mu^i}\,e\indices{_\nu^j}\,\left(\omega\indices{_j_i_\alpha}+\omega\indices{_i_j_\alpha}\right)=0.
\end{equation}

\medskip
The formally introduced gauge field remains an external constraint unless its dynamics is specified via a (``kinetic'') Hamiltonian fixing its vacuum dynamics.
Hence in order to close the system, a free gravity Hamiltonian density $\tilde{\HCd}_{\mathrm{Gr}}$ was added in~\eref{def:actionintegral0}.
However, it is important to stress here that the action integral~\eqref{def:actionintegral0} is generic as it has been {derived} exclusively from the transformation properties of the fields without specifying any involved free field Lagrangians or Hamiltonians!

\section{Computation with the Gasiorowicz parameter}\label{appGasi}
We start by computing the additional contribution in the spin tensor
\begin{align*}
\frac{l}{2}\left(\bar{\psi}\sigma^{\nu\beta}\sigma^{\mu\kappa} \Dderr_\kappa +\Dderl_\kappa \sigma^{\mu\kappa}\sigma^{\nu\beta}\psi\right).
\end{align*}
Performing the contraction with the Levi-Civita tensor and utilizing $\varepsilon_{abcd}\sigma^{bc}=2i\sigma_{ad}\gamma^5=-2\gamma^5(\gamma_a\gamma_d-g_{ad})=2(\gamma_d\gamma_a-g_{da})\gamma^5$ \cite[eq. 4.13]{diracIdentities}, we obtain
\begin{align}
\frac{l}{2}\epsilon_{\xi\nu\beta\mu}\Big(\bar{\psi}& \sigma^{\nu\beta}\sigma^{\mu\kappa}\Dderr_\kappa +\Dderl_\kappa \sigma^{\mu\kappa}\sigma^{\nu\beta}\psi\Big)\nonumber\\
&=\frac{\ell}{2}\bar{\psi}\Big[\gamma^5(\gamma_\xi\gamma_\mu-g_{\xi\mu})\sigma^{\mu\kappa}\Dderr_\kappa+\Dderl_\kappa \sigma^{\kappa\mu}(\gamma_\mu\gamma_\xi-g_{\mu\xi})\gamma^5\Big]\psi\nonumber\\
&=-i\frac{\ell}{2}\bar{\psi}\Big[\gamma^5(\gamma_\xi\gamma_\mu-g_{\xi\mu})(\gamma^\mu\gamma^\kappa-g^{\mu\kappa})\Dderr_\kappa\nonumber\\
&\qquad+\Dderl_\kappa(\gamma^\kappa\gamma^\mu -g^{\mu\kappa})(\gamma_\mu\gamma^\xi-g_{\mu\xi})\gamma^5\Big]\psi\nonumber\\
&=-i\frac{\ell}{2}\bar{\psi}\Big[\gamma^5(2\gamma_\xi\gamma^\kappa+\delta^\kappa_\xi)\Dderr_\kappa +\Dderl_\kappa(2\gamma^\kappa \gamma_\xi +\delta^\kappa_\xi)\gamma^5\Big]\psi\nonumber\\
&=\left(\ell m+\ell^2\frac{\bar{R}}{4}\right)\bar{\psi}\gamma_\xi\gamma^5\psi-i\frac{\ell}{2}\bar{\psi}\left(\gamma^5 \Dderr_\xi +\Dderl_\xi \gamma^5\right)\psi.
\end{align}
In the last step, we substituted the modified Dirac equation (\ref{dirac}) and neglected all torsion-related terms.\par
The additional terms arising from the energy momentum tensor are
\begin{align}
i\frac{\ell}{2}\bar{\psi}\Dderl_\alpha\left(\sigma_\beta{}^\lambda\delta_\nu^\alpha-\sigma_\nu{}^\lambda\delta_\beta^\alpha-\sigma_\beta{}^\alpha\delta_\nu^\lambda+\sigma_\nu{}^\alpha\delta_\beta^\lambda\right)\Dderr_\lambda \psi.
\end{align}
Using $i\sigma_{ab}=\gamma_a\gamma_b-g_{ab}$ and inserting the Dirac equation, we obtain for the first term
\begin{align}
\bar{\psi}\Dderl_\alpha\delta_\nu^\alpha\sigma_\beta{}^\lambda \Dderr_\lambda \psi&=-i\bar{\psi}\Dderl_\nu (\gamma_\beta\gamma^\lambda-\delta^\lambda_\beta)\Dderr_\lambda\psi\nonumber\\
&=-\bar{\psi}\Dderl_\nu \left[\gamma_\beta\left(m+\frac{\ell}{4}\bar{R}\right)-\Dderr_\beta\right]\psi.
\end{align}
The second term follows immediately upon swapping $\beta$ and $\nu$. For the third term, we obtain
\begin{align}
\bar{\psi}\Dderl_\alpha\sigma^\alpha{}_\beta\delta^\lambda_\nu \Dderr_\lambda\psi&=-i\bar{\psi}\Dderl_\alpha(\gamma^\alpha\gamma_\beta-\delta^\alpha_\beta)\Dderr_\nu\psi\nonumber\\
&=\bar{\psi}\left[\left(m+\frac{\ell}{4}\bar{R}\right)\gamma_\beta +\Dderl_\beta\right]\Dderr_\nu\psi.
\end{align}
Upon adding all four expressions we see, that the terms involving two derivatives cancel and the remaining ones can be combined with the ones used in the main derivation. We are thus left with
\begin{align}
T_{\text{D}[\nu\beta]}=\frac{i}{2}&\left(\frac{1}{2}+\ell m+\frac{\ell^2}{4}\bar{R}\right)\bar{\psi}\left(\Dderl_\beta\gamma_\nu-\Dderl_\nu\gamma_\beta+\gamma_\beta\Dderr_\nu-\gamma_\nu\Dderr_\beta\right)\psi.
\end{align}
The following computations are identical to the ones performed in the main paper, so we will not repeat them here. Only the pre factor changes and an additional term $\propto \bar{\nabla}_\mu \bar{R}$ arises.\par
Putting all this together, we arrive at the final expression for the full source without any negligence:
\begin{align}
 g_1 Q_\xi&=\left(\frac{3}{2}+\ell m+\frac{\ell^2}{4}\bar{R}\right) j^A_\xi +i\frac{\ell}{2}\bar{\psi}\left(\gamma^5 \Dderr_\xi+\Dderl_\xi \gamma^5 \right)\psi\nonumber\\
&\quad+\frac{i(1+2\ell m +\ell^2 \bar{R}/2)}{M_\text{P}^2 +g_3}\varepsilon_\xi{}^{\beta\nu\mu}\bar{\psi}\Dderl_\beta\gamma_\nu\Dderr_\mu\psi\nonumber\\
& \quad+i\frac{\ell^2}{8(M_\text{P}^2-g_3)}\varepsilon_\xi{}^{\nu\beta\mu}\left(\bar{\nabla}_\mu \bar{R}\right)\bar{\psi}\left(\Dderl_\beta\gamma_\nu-\gamma_\nu\Dderr_\beta\right)\psi
\end{align}
\section{Derivative of the stress-energy tensor}\label{appTD}
To bring the derivative of the fermionic stress-energy tensor into a more insightful form, we just have to complete the covariant derivatives by inserting appropriate spin connections.
\begin{align}
\nabla_\mu (\bar{\psi}\gamma_\beta \Dderr_\nu \psi)&=\partial_\mu (\bar{\psi}\gamma_\beta \Dderr_\nu \psi)-\Gamma^\alpha{}_{\beta\mu}\bar{\psi}\gamma_\alpha\Dderr_\nu \psi\nonumber-\Gamma^\alpha{}_{\nu\mu}\bar{\psi}\gamma_\beta \Dderr_\alpha \psi\nonumber\\
&=\partial_\mu\bar{\psi}\gamma_\beta\Dderr_\nu\psi +\bar{\psi}\partial_\mu \gamma_\beta\Dderr_\nu\psi +\bar{\psi}\gamma_\beta \partial_\mu (\Dderr_\nu \psi)\nonumber\\
&\quad-\bar{\psi}\Gamma^\alpha{}_{\beta\mu}\gamma_\alpha\Dderr_\nu\psi -\bar{\psi}\gamma_\beta \Gamma^\alpha{}_{\nu\mu}\Dderr_\alpha\psi\nonumber\\
&=\bar{\psi}\Dderl_\mu \gamma_\beta \Dderr_\nu \psi +\bar{\psi}\bm{\omega}_\mu \gamma_\beta \Dderr_\nu \psi +\bar{\psi}\partial_\mu \gamma_\beta \Dderr_\nu \psi\nonumber\\
&\quad +\bar{\psi}\gamma_\beta \Dderr_\mu \Dderr_\nu \psi -\bar{\psi}\gamma_\beta \bm{\omega}_\mu \Dderr_\nu \psi\nonumber\\
&\quad-\bar{\psi}\Gamma^\alpha{}_{\beta\mu}\gamma_\alpha\Dderr_\nu \psi -\bar{\psi}\gamma_\beta \Gamma^\alpha{}_{\nu\mu}\Dderr_\alpha\psi\nonumber\\
&= \bar{\psi}\left(\Dderl_\mu \gamma_\beta\Dderr_\nu -\gamma_\beta \Gamma^\alpha{}_{\nu\mu}\Dderr_\alpha +\gamma_\beta \Dderr_\mu\Dderr_\nu \right)\psi\nonumber\\
&\quad+\bar{\psi}\left(\partial_\mu \gamma_\beta +\bm{\omega}_\mu \gamma_\beta -\gamma_\beta \bm{\omega}_\mu-\Gamma^\alpha{}_{\beta\mu}\gamma_\alpha\right)\Dderr_\nu \psi.
\end{align}
It is easy to show that the bracketed term in the last line combines to the covariant derivative of the tetrad times a gamma matrix, so that it vanishes.
\section{Representations of the Riemann-Cartan tensor}\label{appRep}
We wish to derive the relation between the field strength tensor of the spin covariant derivative resulting from its commutator and the Riemann-Cartan tensor. We start by computing the commutator of the spin covariant derivative, with the Clifford algebra valued gauge field being related to the scalar spin connection through
\begin{align}
\bm{\omega}_\alpha=\frac{i}{4}\omega^i{}_{j\alpha}\sigma^j{}_i.
\end{align}
It is obvious, that the linear terms in the Riemann-Cartan tensor are related to the Clifford valued connection by
\begin{align}
\partial_\alpha \bm{\omega}_\beta=\frac{i}{4}\partial_\alpha\omega^i{}_{j\beta}\sigma^j{}_i.
\end{align}
\indent Concerning the non linear terms, we can make use of the well known identity for the commutator of two sigma matrices \cite[eq. 2.11]{Macfarlane:1966vba}
\begin{align*}
[\sigma^j{}_i,\sigma^l{}_k]=2i(\delta^j_k \sigma_i{}^l -\eta_{ik}\sigma^{jl}+\delta_i{}^l \sigma^j{}_k-\eta^{jl}\sigma_{ik}),
\end{align*}
so that it follows
\begin{allowdisplaybreaks}
\begin{align}
\bm{\omega}_\beta\bm{\omega}_\alpha-\bm{\omega}_\alpha \bm{\omega}_\beta&=-\frac{1}{16}(\omega^i{}_{j\beta}\sigma^j{}_i\omega^k{}_{l\alpha}\sigma^l{}_k-\omega^k{}_{l\alpha}\sigma^l{}_k\omega^i{}_{j\beta}\sigma^j{}_i)\nonumber\\
&=-\frac{1}{16}\omega^i{}_{j\beta}\omega^k{}_{l\alpha}[\sigma^j{}_i,\sigma^l{}_k]\nonumber\\
&=-\frac{i}{8}\omega^i{}_{j\beta}\omega^k{}_{l\alpha}(\delta^j_k \sigma_i{}^l -\eta_{ik}\sigma^{jl}+\delta_i{}^l \sigma^j{}_k-\eta^{jl}\sigma_{ik})\nonumber\\
&=-\frac{i}{8}(\omega^i{}_{j\beta}\omega^j{}_{l\alpha}\sigma_i{}^l-\omega_{kj\beta}\omega^k{}_{l\alpha}\sigma^{jl}\nonumber\\
&\qquad\quad+\omega^i{}_{j\beta}\omega^k{}_{i\alpha}\sigma^j{}_k-\omega^i{}_{j\beta}\omega^{kj}{}_\alpha\sigma_{ik})\nonumber\\*
&=\phantom{-}\frac{i}{4}(\omega^i{}_{n\beta}\omega^n{}_{j\alpha}-\omega^i{}_{n\alpha}\omega^n{}_{j\beta})\sigma^j{}_i.
\end{align}
\end{allowdisplaybreaks}
\indent It is now obvious, that
\begin{align}
\bm{R}_{\alpha\beta}&:=[\Dderr_\alpha,\Dderr_\beta]=[\Dderl_\beta,\Dderl_\alpha]\nonumber\\
&\phantom{:}=\partial_\alpha \bm{\omega}_\beta-\partial_\beta\bm{\omega}_\alpha+\bm{\omega}_\alpha\bm{\omega}_\beta-\bm{\omega}_\beta\bm{\omega}_\alpha\nonumber\\
&\phantom{:}=\frac{i}{4}\left(\partial_\alpha\omega^i{}_{j\beta}-\partial_\beta\omega^i{}_{j\alpha}+\omega^i{}_{n\alpha}\omega^n{}_{j\beta}-\omega^i{}_{n\beta}\omega^n{}_{j\alpha}\right)\sigma^j{}_i\nonumber\\
&\phantom{:}=\frac{i}{4}R^i{}_{j\alpha\beta}\sigma^j{}_i.
\end{align}

\end{document}